\documentclass[11pt]{article}
  \usepackage{amssymb}
   \usepackage{graphicx}

\def\al {\alpha}

\def\ba{\begin{eqnarray}}
\def\bam{\begin{array}}

\def\be{\begin{equation}}

\def\bi{\bibitem}

\def\bt {\beta}

\def\B {\overline}

\def\de{\delta}

\def\ea{\end{eqnarray}} 
\def\ee{\end{equation}}

\def \EE{{\cal E}}

\def\fr{\frac}

 \def\ga{\gamma}

\def\ha{\frac{1}{2}~}

\def\in{\infty}

\def\ka{\kappa}

\def\lb{\label}

\def\na{\nabla}

\def\nn{\nonumber}

\def\td{\tilde}

\def\ts{\textstyle}

\def\tt{\tilde t}

\def\vf{\varphi }

\def\ze{\zeta}

\def\1{{\it one}}

\def\2{{\ts{\ha}\!}}

\def\3 {\ts{\frac{1}{3}\!}}
\def\4{\ts{\fr{1}{4}\!}}

\title{The Torqued Cylinder and Levi-Civita's metric}
\author{ Donald Lynden-Bell$^{1,2}$\thanks{email:dlb@ast.cam.ac.uk}
\\  {\it$^1$  Institute of Astronomy, Madingley Road, Cambridge CB3 0HA,United Kingdom }
\\ {\it$^2$ The Racah Institute of Physics, Edmond Safra Campus, Jerusalem, Israel}}
\date{\today}                                           
\begin{document}
\maketitle
Abstract\\
When a static cylindrical system is subjected to equal and opposite torques top and bottom
it transports angular momentum along its axis. The external metric of this static  system can 
be transformed to Levi-Civita's form by using helical coordinates. This gives the external 
metric of a static cylinder three dimensionless parameters corresponding to the mass per unit length, the total stress along the cylinder, and the total torque.\\
	The external vacuum metric of a spherical system is characterised by its mass alone. How many parameters characterise the external metric of a general stationary cylindrical system? Leaving aside the radius of the cylinder which defines the scale we find that there are five parameters, the three above mentioned, to which should be added the momentum along the cylinder per unit length and the angular momentum per unit length. We show how to transform Levi-Civita's one parameter metric to include all five.
 \section{Introduction}
 In its standard dimensionless form with $c=1$ Levi-Civita's metric for the external space-time of a cylindrical system is
 \be
 d\B s^2=\B\rho^{2m}d\B t^2-\B\rho^{~-2m}[\B\rho^{2m^2}(d\B\rho^2+d\B z^2)+\B\rho^2d\B\vf^2].\lb{11}
 \ee
 With coordinates in the order $\B t,\B\rho,\B\vf,\B z$  three independent Killing vectors are
 ~~$\B\xi^\mu =(1,0,0,0),\\ \B\eta^\mu=(0,0,1,0),~~\B\ze^\mu=(0,0,0,1).$ Given the suggestive way in which the metric is written it is natural to suppose that $\B t$ should be interpreted as time, $\B z$ as coordinate height up the axis and $\B\vf$ as the azimuthal angle about the axis with the range $0\le\B \vf< 2 \pi$. However such an interpretation misses the amazing fecundity of this metric. Any linear combination of Killing vectors with constant coefficients gives another
 Killing vector so the true time might be such a linear combination of the above coordinates
 and likewise the true $z$ coordinate and the true azimuth. Even for static cylinders the
 assumed range of $\B\vf$ is unnecessarily restrictive and for cylindrical shells it gives only cylinders with no net longitudinal stress.  Bi\v{c}\'{a}k and \v{Z}ofka \cite{BZ} have emphasised this and removed the problem by introducing $\td\vf=C\B\vf$ with  $\td\vf$ having the desired range. They call $C$ the conicity since, were this metric to hold down to the axis, it would have a conical singularity there of magnitude $C-1$. Putting $\B b \B t=\tt,~~\B\rho=\B R/\B b,~~\B b \B z=\B Z$ and $ds=\B bd\B s$ their metric is
 \be
 ds^2= \B\rho^{2m}d\tt^2-\B\rho^{~-2m}[\B\rho^{2m^2}(d\B R^2+d\B Z^2)+\B R^2d\td\vf^2/C^2];\lb{12}
 \ee
  it has two dimensionless parameters, Levi-Civita's $m$ and the conicity. For cylindrical shells of radius $\B b$ the rest mass per unit length is $\mu$, where $G\mu/c^2=\4[1-(1-m)^2/C]$ and the total stress along $z$ is $\Pi$ where $\ka \Pi=2\pi(C^{-1}-1);~~~\ka=8\pi G/c^4$\cite{BZ}.  The implications of this for the external field of cylinders full of matter is detailed in \cite{BLSZ}.  Ji\v{r}\'{i} Bi\v{c}\'{a}k pointed out to me that for filled cylinders the vertical stress was not a function of $C$ alone.
This led me to the interesting interpretation of the meaning of $C$ discussed below. 
 Levi Civita's mass parameter $m$ while mathematically convenient was never a good notation as even in the classical limit the mass per unit length is not $m$ but $m/2$. Once $C$ has been introduced the rest mass per unit length is not a function of $m$ alone but involves $C$ too. However the gravity field, defined as minus the force per unit mass required to keep a particle at rest on the time-like Killing vector, is still $\B\EE_{\B R}=-d\ln\xi/d\B R=-m/\B R$ (where $\xi^2=g_{00}$). This may be compared with the classical result for cylinders with mass per unit length $\mu,$ viz. $-2G\mu/R$. The Bi\v{c}\'{a}k and \v{Z}ofka, metric (\ref{12}) is no longer in Weyl's form but may be put in that form by a rescaling  $\td R=\B R/C,~~\td Z=\B Z/C$, so $\B\rho=\td R/b$ where $b=C\B b$
\be
ds^2= \B\rho^{2m}d\td t^2-\B\rho^{~-2m}[C^2\B\rho^{2m^2}(d\td R^2+d\td Z^2)+\td R^2d\td\vf^{~2}];\lb{13}
\ee
 [Putting $C=1$ one recovers Levi-Civita's dimensional form.]
 Inside a filled cylinder the metric can be written with $\psi(0)=0$,
 \be
 ds^2=e^{-2\psi}d\tt^2-e^{2\psi}(e^{2a}d\td R^2+\td R^2d\td\vf^2+e^{2\ze}d\td Z^2).
 \ee
 Writing $'$ for an $\td R$ derivative, the mixed $zz$ component of Einstein's equations reads
 \be
 e^{-2(\psi+a)}(\psi'^{~2}-a'/\td R)=\ka p_z^z.
 \ee
 Since $a(0)=0$ to avoid a singularity on the axis, this can be integrated in the form
 \be
 2\pi [e^{-a}-1]/\ka=\int_0^{\B R} ( p_z^z-|{\bf grad}\psi |^2/\ka)dS;~~~dS=(2\pi e^\psi \td R)e^{(\psi+a)}d\td R.\lb{16}
 \ee
  At the edge of the distribution  $\td R=b$ and there $e^{a(b)}=C$ so the integral is $2\pi(C^{-1}-1)/\ka$, a formula which was already found for cylindrical shells by \cite{BZ}. For shells there is no contribution from the  $|{\bf grad}\psi|^2/\ka$ term inside as the potential is constant there.
 Outside any cylinder $\td R>b$ the contribution to the integral can be explicitly evaluated since $\psi=-m\ln(\td R/b)$ and $e^a=C(\td R/b)^{m^2}$;
 \be
 \int_b^\in-(|{\bf grad}\psi |^2/\ka) dS=-2 \pi/(C\ka)\lb{17}.
 \ee
 To interpret these formulae we first turn to the Classical formulae of Morgan and Bondi \cite{MB} for the stress tensor of the Newtonian gravitational field $[\na\psi\na\psi-\delta|\na\psi|^2/2]/(4\pi G)$. Here $\de$ is the unit tensor. This has been used to calculate the gravitational torques due to the spiral structure of galaxies \cite{LKl}. The physics of this stress tensor is that the lines of gravitational force carry a pressure $|{\bf grad}\psi|^2/(8\pi G) $ along those lines and a tension of the same magnitude perpendicular to the lines, just the opposite of the Maxwell stresses in electromagnetism.
Evidently in the nearly classical case the formula (\ref{17}) tells us that $2\pi/(\ka C)$ is to be interpreted as the total downward force on the upper part due to the gravitational stresses outside the cylinder, while the formula (\ref{16}) gives the upward stress within a radius $\td R<b$ as the difference between the integral of the $p_z^z$ component of the pressure tensor and the gravitational stresses within $\td R$. It is surprising that a stress tensor imported from Newtonian gravitation seems to work in the alien world of strong field Relativity. While this may hint at the the existence of a stress-energy tensor for the relativistic gravitational field, we found it somewhat strange that the total downward stress across the cut on the upper part of a cylinder is just $2\pi/\ka$, independent of its mass or its material stress.\\
 Where possible, we consider it desirable to use the length of the azimuthal  Killing vector as the radial coordinate so we write $R=\td R(\td R/b)^{-m}$. Then putting $n=1/(1-m) ; \td z=\td Z/C;~\rho=R/b$ the metric becomes for $m\ne1$
 \be
 ds^2= \rho^{2nm}d\td t^2-[n^2 C^2\rho^{2nm^2}d R^2+R^2d\td\vf^2+\rho^{-2m}d\td z^2].\lb{18}
\ee
 	However we find that even after the introduction of $C$, the metric is still too restrictive in that it can not describe the static external space of a torqued cylinder (in English parlance one that carries a couple). The external space-time of such a system needs three dimensionless parameters related to the mass per unit length, the total stress and the total torque. We show here that the external metric of a static torqued cylinder can be described by Levi-Civita's metric but in spatially helical coordinates. It is the new $\vf$ that is the azimuthal angle around the cylinder with the normal range $0,2\pi$. Levi-Civita's coordinates run helically around our cylinder.
Our resultant static metric has three dimensionless parameters, $m,C,\al$ as desired.
Whereas this completes the characterisation of static external metrics it by no means exhausts the amazing fecundity of Levi-Civita's metric. Not only can the spatial Killing vectors be recombined as above but they can be combined with the time-like Killing vector. Many years ago Frehland \cite{Fr} discovered that the external metric of a rotating cylinder could be rewritten locally in static form, and since the Einstein equations are local, only the boundary conditions give any non-locality. Stachel \cite{St} then pointed out the global role played by the assumed topology of the coordinates and made the percipient remark that the interference
of light beams passing on each side of a rotating cylinder would be affected by the rotation
of the cylinder even when its radial gravity was held fixed. This gives a purely classical analogue of the Aharanov-Bohm effect in quantum mechanics. Frehland uses a combination of Levi-Civita time and azimuth for his new time.
More obviously one may use a combination of those with the $\td z$ coordinate to get the 
metric of a cylinder that is both moving along its axis and rotating around it.
 \section{A cylinder transporting linear and angular momentum}
 To hold itself up against its own gravity, a cylindrical shell must have a pressure (or stress) tensor
 with a sufficient azimuthal component pushing around the cylinder. If the cylinder is under more  stress parallel to its axis then its principal axes of stress will be a large stress in that direction and a smaller stress azimuthally. Such a cylinder transports the $z$ component of linear momentum upwards which is the same as transporting negative $z$-momentum downwards; these are not separate conservation laws of upward and downward momentum, they are restatements of the one law of conservation of the $z$ component of momentum. As I have encountered several good physicists who find difficulty with the concept that a static stressed body is carrying a flux of momentum, I here give an example where this occurs. A classical rigid vertical rod is hit by two equal balls each moving along the axis of the rod. The upper ball bounces off the rod's top the lower ball bounces simultaneously off the bottom. The rod never moves but at the instant when the balls hit, it carries a stress. Now consider the flow of $z$- momentum from the lower ball. It passes through the rod which {\it when it is stressed} carries that flux of upward momentum to the upper ball. That statement is the one complete conservation statement. There is a flux of $z$-momentum upward through the unmoving stressed rod. The situation is perfectly symmetrical as is easily seen by considering the downward flux of downward momentum through the stressed rod to the lower ball. The difficulty lies in understanding that in this case the downward flux of downward momentum is the very same thing as the stress which is also the upward flux of upward momentum.\\
 We may represent stress tensors at each point on the cylinder by drawing ellipses on it whose principal axes are proportional to the principal axes of stress. If now the cylinder is carrying a torque (or couple), as might be generated by holding the bottom of a solid shell fixed and twisting the top, then the stress tensor ellipses will be tilted as shown in figure 1. The larger pressure along the longer axes of these tilted ellipses gives a torque that carries angular momentum upwards. We use the usual $\vf,z$ coordinate lines on our shell and see that principal axes of stress are tilted with respect to them. Flat space is the the only static cylindrical empty solution of Einstein's equations that is regular on the axis.  We  take the flat internal metric to be the usual $ds^2=dt^2-(dR^2 +R^2d\vf^2+dz^2)$. This must fit the exterior metric on $R=b$ in such a way that it gives a pressure tensor whose principal axes make helices on the shell. The space part of the pressure tensor has components $P_{(R)(R)},P_{(R)(\vf)},P_{(\vf)(\vf)}$ where the brackets on indeces denote frame components in the local tetrad. The time-time component is $\sigma$ and  the radial component of the contracted Bianchi identity is the only constraint. Thus there are three independent components in the surface stress-energy tensor. To fit them the external metric must have three parameters corresponding to the mass per unit length, the momentum current up the cylindrical shell and the angular momentum current up it. The original Levi-Civita metric has only one parameter, $m$, but in the form given in equation (\ref{12}) it has two, $m,C$; now a third is needed in its more general form. 
 \begin{figure}[htbp]
\begin{center}
\includegraphics[width=8cm]{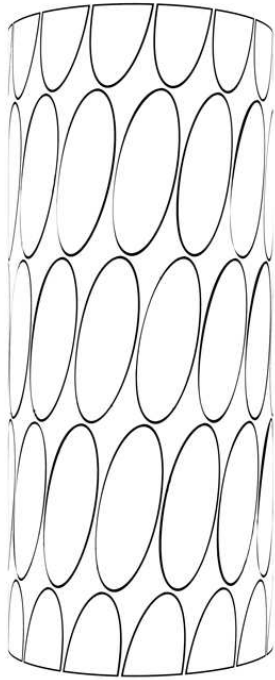}
\caption{The inclined pressure ellipses of the torqued stressed cylindrical shell.}
\label{fig1}
\end{center}
\end{figure}
 The $z\rightarrow-z$ symmetry of all the metrics mentioned so far is not obeyed by the stress tensor of the torqued shell but it does have the $(z,\vf)\rightarrow(-z,-\vf)$ symmetry. 
 The empty metric with that symmetry can be found by plotting $\td z/b$ against $\B \vf$ and  making the transformation $~\td\vf=(\cos\al~\vf-\sin\al~z/b),~~\td z/b=(\cos\al~z/b+\sin\al~\vf), $ in the  metric (\ref{18}).  This transformation is orthogonal on the cylinder $R=b$.  Putting  $l_1=\rho^{-2m}\cos^2\al+\rho^2\sin^2\al$ and  $l_2=\rho^{-2m}\sin^2\al+\rho^2\cos^2\al $ the resulting metric was first given in \cite{LBK}
 \ba
 &&ds^2=\rho^{2nm}d\td t^2-[ n^2 C^2 \rho^{2nm^2} dR^2+ l_2 b^2 d\vf^2 + ~\nn\\
&&~~~~~~~~+[b\sin2\al(\rho^{-2m}-\rho^2)d\vf dz+l_1 dz^2]
=\xi^2 dt^2-\ga_{kl}dx^k dx^l.
\ea
 We shall refer to $\ga_{kl},k,l=1,2,3$ as the gamma metric of space as opposed to $ds^2 $ the metric of space-time which now has three parameters $m,C,\al$. It reduces to the internal metric on the surface $R=b,\rho=1$ where $dR=0,~l_1=l_2=1$. 
While $z$ and $\vf$ are orthogonal on the cylinder, at larger $R$ the cross term shows they are not. Nevertheless the gamma metric in the square brackets is everywhere positive definite.
We now  fit this external metric to the stress-energy tensor of a shell carrying both torque and upward stress. Employing the variation of Israel's technique used earlier \
\cite{LK} the stress energy tensor integrated across the shell is given by
\ba
&&\sigma=(\ka b)^{-1}[1-(1-m)^2/C],~~~~~~~\nn\\
&&P_{\vf\vf}=\ka^{-1}(b/C)[m^2\cos^2\al+\sin^2\al],~~~~~~~~~~\nn\\
&&P_{\vf z}=\ka^{-1}(1-m^2)C^{-1}\sin\al \cos\al,~~~~~~~~~~\nn\\
&&P_{zz}=(\ka b)^{-1}[C^{-1}\cos^2\al-1+m^2C^{-1}\sin^2\al].~
\ea
These  agree with the results  of \cite{BZ} when there is no torque.\\
The total flux of  momentum up the cylinder is  $2\pi b P_{zz}$. and the flux of angular momentum is $\dot L=2\pi bP_{\vf z}$.
A mathematician might complain that all we have done is to change coordinates in  Levi-Civita's metric but that is a mathematician's viewpoint taking no regard to the interpretation and the ranges of the variables. The physical cylinder could have the coordinates $\vf,z$ drawn on it and these agree with the metric induced upon it from both the internal and the external metric. The external coordinates $\B t,R,\vf,z$ still have the Killing symmetries in $\B t,\vf,z$ but $R^2 d\vf/ds$ is not the specific angular momentum of a test particle in orbit  Instead the specific angular momentum is $h=l_2d\td\vf/ds+b\sin\al\cos\al [\rho^{-2m}-\rho^2]dz/ds$ and the z-momentum is $p=l_1dz/ds+b\sin\al\cos\al[\rho^{-2m}-\rho^2]d\vf/ds$.   If we return to Levi-Civita's metric in the form (1.4) with coordinates $\td\vf, \td z$  we get back to our usual formulae $\B h=R^2d\td \vf/ds:~ \B p=\rho^{-2m}d\td z/ds$ but those coordinates wind in helices around our cylinder and $\B h,~\B p$ though conserved are not the specific angular momentum and momentum about and along the  axis of our cylinder. However writing these quantities in our new coordinates $\B h=R^2[\cos\al~d\vf/ds-\sin\al~(dz/ds)/b]$;\\
 $\B p=\rho^{-2m}[\cos\al~ dz/ds+b\sin\al~ d\vf/ds]$. Solving these for $d\vf/ds,dz/ds$ we find,
\ba
d\vf/ds=\cos\al~\B h/R^2+\sin\al~\B p~\rho^{2m}/b;\nn\\~~~ dz/ds=\cos\al~\B p~\rho^{2m}-\sin\al~b\B h/R^2.
\lb{211}
\ea
Expressing the new linear and angular momenta in terms of the old,
\be
p=-(\B h/b)\sin\al+\B p\cos\al;~~~h=\B h\cos\al+b\B p\sin\al.
\lb{212}
\ee
 \begin{figure}[htbp]
 \begin{center}
\includegraphics[width=8cm]{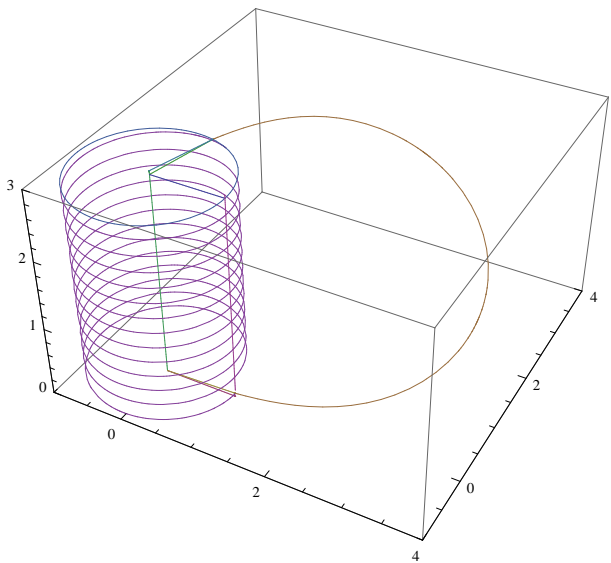}
\caption{ A geodesic with no angular momentum starts from the origin moving outward and upward in the xz-plane;  it emerges through the torqued cylinder still in that plane but the helical distortion of the space outside causes it to move in azimuth as well as radius and height. When it re-enters the cylinder it again moves in a plane through the axis. We chose the shear $\tan \al=0.5$ for this illustration.}
\label{fig2}
\end{center}
\end{figure}
What  gravitational effects are produced by a cylinder that carries torque? Firstly since $\xi$ depends only on $R$, its gravity is still radially directed to the axis. The geodesics inside the cylinder with zero angular momentum lie in planes $\vf=const$ but if we start a particle moving outward and upward from the cylinder's surface then initially $d\vf/ds=0$ and from equation (\ref{211})  $\B h/b=-\B p\tan\al$, so outside the cylinder it obeys $d\vf/ds=(\B p/b)\rho^{-2}(\rho^{2m+2}-1)\sin\al$, which maintains its sign since outside $\rho>1$. When it  returns to the cylinder it will do so at another azimuth $\vf$ but at that moment it will again be moving in an $R,z$ plane since $\rho=1$,there. Thus the zero angular momentum geodesics wind around the cylinder whenever $\B p \ne 0$. Exactly radial geodesics are the only ones that do not wind. The forces on particles forced to move along the Killing vectors of this space are all radial since they can be put in the form (\ref{212}) and since all metric coefficients have only an $R$-dependence the z-Killing vector $\ze$ will have just that dependence. ( It will of course depend on the constants that define which Killing vector we are forcing our particle to move along.) However there are forces in the $\vf$ direction for trajectories that move in both $R$ and $z$. 
The geodesics are fully integrable as is readily seen when we add the energy equation $\rho^{2nm}dt/ds=E$ to the integrals already found in (\ref{211}). Of course one must use the metric equation divided by $ds^2$ too. Readers will find that their dynamical intuition to be somewhat challenged by the motions in this metric.\\
\section{ The general stationary exterior cylindrical metric}
We now start from the metric (\ref{18}) but use $b\td\vf$ as our azimuthal coordinate in place of  $\td\vf$
\be
 ds^2= \rho^{2nm}d\td t^2-[n^2 C^2\rho^{2nm^2}d R^2+R^2d(b\tilde\vf)^2+\rho^{-2m}d\td z^2].
 \ee
 On the surface of our cylinder $R=b$ it reduces to
\be
ds^2=d\tilde t^2-(b^2d\tilde\vf^2+d\tilde z^2).
\ee
This metric is flat and its form  is invariant under rotations in $b\tilde\vf-\td z$ as exploited above
for the torqued cylinder. However it is also invariant  under Lorentz transformations both 
in  $\td t,\td z$ and in $\td t,b\tilde\vf$ and also in any combination of them. Such transformations 
with constant coefficients when applied not just to this $R=b$ surface but to all of space-time give us the metric generated by a rotating torqued cylinder under vertical strain that is also in motion along its axis.\\
To make the argument easy to follow we shall transform the metric by first making a Lorentz transformation at speed $v$ at a general angle $\bt$ to the $\td z$ axis in the $\td z,b\tilde\vf$ plane. Such a transformation leaves the plane  invariant but changes the coordinates. We shall follow this transformation by a relatively small rotation through an angle $\al$ corresponding to the torque on the cylinder considered in the last section. Finally we announce that our new $t$ is to be considered as the time, our new $z$ is to be considered as the coordinate along the axis, and our new $\vf$ is to be the angle about the axis and we impose the appropriate periodicity $0\le\vf< 2\pi$. Although these transformations can be applied to the exterior metric of any cylindrical system we find it helpful to have a definite system in mind and we choose the simplest possible the cylindrical shell of radius $b$.
The details of a Lorentz transformation followed by a rotation are straightforward so to save 
the reader from tedium we shall give the net result; the old tilde coordinates are related to the new unadorned ones by the linear transformation\\
\ba
\td t=a_{00}t+a_{02}b\phi+a_{03}z,\nn\\
b\td\vf=a_{20}t+a_{22}b\vf+a_{23}z,\nn\\
\td z=a_{30}t+a_{32}b\vf+a_{33}z.
\ea
The constant coefficients are all determined by $v,\bt,$ and $\al$ and take the rather lengthy form 
\ba
&&a_{00}=\ga=(1-v^2)^{-\2}; ~~a_{02}=v\ga\sin(\bt-\al);~~a_{03}=v\ga\cos(\bt-\al);\nn\\
&&a_{20}=v\ga\sin\bt;~~a_{22}=[(\ga-1)\sin^2\bt+1]\cos\al-(\ga-1)\sin\bt\cos\bt\sin\al;\nn\\
&&a_{23}=[(\ga-1)\sin^2\bt+1]\sin\al+(\ga-1)\sin\bt\cos\bt\cos\al;\nn\\
&&a_{30}=v\ga\cos\bt;~~a_{32}=-[(\ga-1)\cos^2\bt+1]\sin\al+(\ga-1)\sin\bt\cos\bt\cos\al;\nn\\
&&a_{33}=[(\ga-1)\cos^2\bt+1]\cos\al+(\ga-1)\sin\bt\cos\bt\sin\al.
\ea
The metric now has the five independent constants $m,C,v,\al,\bt$ as well as the overall scale constant $b$.
We write it in the form
\be
ds^2=\xi^2(dt+A_2 b d\vf+A_3dz)^2-\ga_{kl}dx^kdx^l,
\ee
where
\ba
&&\xi^2=\rho^{2nm}a_{00}^2-\rho^2 a_{20}^2-\rho^{-2m}a_{30}^2, \nn\\
&&A_2 = (\rho^{2nm}a_{00}a_{02}-\rho^2 a_{20}a_{22}-\rho^{-2m}a_{30}a_{32})/\xi^2, \nn\\
&&A_3 = (\rho^{2nm}a_{00}a_{03}-\rho^2 a_{20}a_{23}-\rho^{-2m}a_{30}a_{33})/\xi^2,\nn\\
&&\ga_{11}=n^2C^2\rho^{2m^2};~~\ga_{12}=\ga_{13}=\ga_{21}=\ga_{31}=0,\nn\\
&&\ga_{22}=(-\rho^{2nm}a_{02}^2+\rho^2 a_{22}^2+\rho^{-2m}a_{32}^2)+\xi^2A_2^2,\nn\\
&&\ga_{23}=\ga_{32}=(-\rho^{2nm}a_{02}a_{03}+\rho^2a_{22}a_{23}+\rho^{-2m}a_{32}a_{33})+\xi^2A_2A_3,\nn\\
&&\ga_{33}=(-\rho^{2nm}a_{03}^2+\rho^2 a_{23}^2+\rho^{-2m}a_{33}^2)+\xi^2A_3^2.
\ea
When there is neither torque nor motion along the axis $\al=0,\bt=\pi/2$ so these formulae
are greatly simplified with $a_{03}=a_{23}=a_{30}=a_{32}=A_3=0$.\\
The standard form of the rotating metric is \cite{Ste}
\ba
ds^2=\rho X(dt+Ad\phi)^2-\rho^{\2(n^2-1)}(d\rho^2+dz^2)-X^{-1}\rho d\vf^2;\nn\\
 X=a_1\rho^n+a_2\rho^{-n};~~~A=\frac{b_1\rho^n}{na_2X}+b_2;~~~b_1^2=-n^2a_1 a_2;\lb{319}
\ea
Since at any radius all the metric coefficients are constant it follows that at any one radius $b$, the induced metric can be written as a 2+1 Minkowski metric after taking $T=t+A(b)\vf$ and scaling the coordinates appropriately. Thus the above analysis applies. However the proof that (\ref{319})
can be reduced to (\ref{11}) is more easily accomplished directly.
Making the transformation $t=b_{00}\td t+b_{02}\td\vf;~~\vf=b_{20}\td t+b_{22}\td \vf$,
we find no $d\td t d\td\vf$ term if for all $\rho$
\be
(b_{00}X+A X b_{20})(b_{02}x+A X b_{22})=b_{20}b_{22}.\lb{320}
\ee
This is satisfied if firstly $ b_{00}+b_{20}b_2=0$, which ensures that the first factor is $\frac{b_1 b_{20}}{n a_2} \rho^n$ and secondly $(b_{02}+b_2 b_{22})a_1+\frac{b_1 b_{22}}{n a_2}=0$ which ensures that the second factor is $ \frac{-b_1 b_{22}}{n a_1}\rho^{-n}$, so that (\ref{320}) is satisfied thanks to (\ref{319}). We notice that we are still free to set $b_{22}^2=a_1$ and $b_{20}^2=-a_2>0$. Writing $2m=n+1$ the metric is reduced to (\ref{11}). This demonstrates Frehland"s point that locally the rotating metrics are of Levi-Civita's form.
Gravitational forces due to these metrics are discussed  in \cite{LBK}
\section{Acknowledgements}
Figures 1 an 2 are taken from \cite{LBK} where they are figures 4 and  6 on pages 7 and 8 of the 
paper published in MNRAS.
This work was completed while visiting the Hebrew University and Profs Tsvi Piran and Joseph Katz whose support is much appreciated. I thank Prof Bi\v c\'ak for his comments on the content. One of the referees was exceptionally thorough and found a number of sign errors which have been corrected thanks to his assiduity.

\end{document}